\documentclass[useAMS,usenatbib,usegraphicx]{mn2e}

\usepackage{amsmath, hyperref, amssymb, color, latexsym, graphicx, amsfonts}
\usepackage{multirow}
\usepackage{xspace}
\usepackage{aas_macros}
\usepackage{textcomp}
\usepackage{subfigure}
\usepackage{times}
\usepackage{bm}

\newcommand{\gws}{gravitational waves\xspace}
\newcommand{\gw}{gravitational wave\xspace}
\newcommand{\ee}[1]{\!\times\!10^{#1}}
\newcommand{\pt}{{\it PT}\xspace}
\newcommand{\et}{{\it ET}\xspace}
\newcommand{\pts}{{\it PT}s\xspace}
\newcommand{\ets}{{\it ET}s\xspace}
\newcommand{\ra}[4]{\ensuremath{\alpha = {#1}^{\rmn{h}} {#2}^{\rmn{m}}
{#3}\fs{#4}}}
\newcommand{\dec}[4]{\ensuremath{\delta = {#1}\degr {#2}\arcmin
{#3}\farcs{#4}}}

\title[Extending burst searches with PTAs]{Extending \gw burst searches with
pulsar timing arrays}

\author[M.~Pitkin]{Matthew Pitkin$^1$\thanks{matthew.pitkin@glasgow.ac.uk} \\
$^1$SUPA, School of Physics and Astronomy, University of Glasgow,
University Avenue, Glasgow, G12 8QQ, UK}
\date{}

\pagerange{\pageref{firstpage}--\pageref{lastpage}} \pubyear{2012}

\begin{document}

\maketitle

\begin{abstract}
Pulsar timing arrays (PTAs) are being used to search for very low frequency
\gws. A \gw signal appears in pulsar timing residuals through two components:
one independent of and one dependent on the pulsar's distance, called the
`Earth term' (\et) and `pulsar term' (\pt), respectively. The signal of a
burst (or transient) \gw source in pulsars' residuals will in general have the
Earth and pulsar terms separated by times of the order of the time of flight
from the pulsar to the Earth. Therefore, both terms are not observable over a
realistic observation span, but the \ets observed in many pulsars should be
correlated. We show that pairs (or more) of pulsars can be aligned in such a way
that the \pts caused by a source at certain sky locations can arrive at Earth
within a time window short enough to be captured during a realistic observation
span. We find that for the pulsars within the International Pulsar Timing Array
(IPTA) $\sim 67$ per cent of the sky produces such alignments for pulsars terms
separated by less than 10 years. We compare estimates of the source event rate
that would be required to observe one signal in the IPTA if searching for the
correlated \ets, or in searching via the \pts, and find that event rates would
need to be about two orders of magnitude higher to observe an event with the
\pts than the \ets. We also find that an array of hundreds of thousands of
pulsars would be required to achieve similar numbers of observable events in \pt
or \et searches. This disfavours \pts being used for all-sky searches, but they
could potentially be used target specific sources and be complementary to \et
only searches.
\end{abstract}


\section{Introduction}
In the late 1970s it was first suggested that precision timing of
pulsars could be used to detect very low frequency \gws \citep{Sazhin:1978,
Detweiler:1979}. This lead to early searches for a cosmological stochastic
background of nanoHz \gws \citep[e.g.][]{Hellings:1983, Romani:1983, Davis:1985,
Rawley:1987, Stinebring:1990, Kaspi:1994}. The first attempt to construct an
array of pulsars for use as a \gw detector (amongst other applications) was that
of \citet{Foster:1990}. More recently several groups around the world (the
Parkes Pulsar Timing Array \citep{Verbiest:2010}, the European Pulsar Timing
Array \citep{Ferdman:2010} and the North American Nanohertz Observatory for
Gravitational Waves \citep{Jenet:2009, Demorest:2012}) have worked to set up and
perform precision timing of a selection of stable millisecond pulsars. The aim
of these is to detect low frequency \gws from objects such as supermassive
binary black holes (SBBH). These are now being combined into a concerted
world-wide effort to form an International Pulsar Timing Array (IPTA)
\citep{Hobbs:2010}. 

Initial \gw searches using pulsar timing focused on looking for a cosmological
stochastic background. Following the theoretical work of
\citet{Jaffe:2003} and \citet{Wyithe:2003} the focus has more recently shifted
to finding a stochastic background from multiple SBBHs. \citet{Lommen:2001}
performed the first searches for individual quasi-monochromatic SBBH sources
\citep[with the more recent theoretical work of][providing more impetus for
this]{Sesana:2009}, with this method being used to rule out a putative
electromagnetic observation of such a system in the radio galaxy 3C\,66B
\citep{Jenet:2004}. Now there are many proposed methods to detect stochastic
sources \citep[e.g.][]{McHugh:1996, Jenet:2005, Anholm:2009, Haasteren:2009},
and several recent searches have been performed providing limits on their
emission \citep[e.g.][]{Jenet:2006, Yardley:2011, Haasteren:2011,
Demorest:2012}. There are now also many proposed methods to search for
individual quasi-monochromatic (or continuous) sources
\citep[e.g.][]{Corbin:2010, Yardley:2010, Lee:2011, Babak:2012, Ellis:2012,
Ellis:2012b}, and even individual short duration transient (or burst) sources
\citep[e.g.][]{Haasteren:2010, Pshirkov:2010, Finn:2010, Cordes:2012}. A review
of the many \gw search avenues currently being explored can be found in
\citet{Lommen:2011}.

Examples of burst sources could be the final inspiral or parabolic encounters of
SBBHs \citep{Finn:2010}, or cosmic string cusps \citep{Leblond:2009,
Binetruy:2009, Key:2009}. Unlike the bursts searched for in ground-based \gw
searches \citep[e.g.][]{Abbott:2009}, which are generally classed as events
lasting of the order of milliseconds--seconds, these {\it bursts} would last
from months to years, but importantly they are transient rather than continuous
signals.

\subsection{Signals in pulsar timing arrays}\label{sec:ptasignal}
Pulsar observers measure the time of arrival (TOA) of pulses, which can be
thought of as ticks of a clock. Timing residuals are the difference between the
observed TOAs and a best fit model of the time of arrival that is dependent on
many parameters such as the pulsar frequency, frequency derivatives, sky
position and any binary system parameters if appropriate. Any unmodelled
components, such as a potential \gw signal, would remain in the residuals.

A \gw signal appears in pulsar timing residuals through two terms:
an `Earth term' and a `pulsar term', which we shall refer to as \et and \pt
from now on. The \et is independent of the pulsar distance and the \pt is
delayed in time from the \et by an amount proportional to the pulsar
distance. This `two-pulse' response is more generally the case for any signal
in a single-arm one-way \gw detector\footnote{Note that for ground-based \gw
detectors, where the wavelength of the \gw is far longer than the arm length,
the `two-pulse' structure and frequency dependence disappear from the detector
response \citep[e.g.][]{Finn:2009}.} e.g.\ \citet{Estabrook:1975} and
\citet{Detweiler:1979}. In this case the pulsar and the receiver on the Earth
represent the ends of the arms of the detector. A signal in the \et will be
simultaneous in all observed pulsar residuals, whereas the \pt will be delayed
by
\begin{equation}\label{eq:deltat}
\Delta{}t = (1 + \hat{\bm{k}}\cdot\hat{\bm{n}})d/c,
\end{equation}
where $\hat{\bm{k}}$ is a unit vector along the \gw propagation direction
(pointing from the source to the Earth), $\hat{\bm{n}}$ is a unit vector
pointing along the Earth-pulsar line-of-sight, and $d$ is the distance to the
pulsar. If we define a plane perpendicular to $\hat{\bm{k}}$ such that the Earth
is at the origin and $\hat{\bm{k}} = \{0, 0, -1\}$, and $\hat{\bm{n}} =
\{\sin{\theta}\cos{\phi}, \sin{\theta}\sin{\phi}, \cos{\theta}\}$, where
$\theta$ is the polar angle between the source and pulsar, and $\phi$ is the
azimuthal angle from some arbitrary point (which for the moment does not
matter), then
\begin{equation}
\Delta{}t = (1 - \cos{\theta})d/c.
\end{equation}
As stated in \citet{Finn:2010}, unless the pulsar and source are closely aligned
on the sky, this time delay will be large. So, for a transient burst source if
the \et is observed then the \pt will not appear in the residuals over the
typical period of pulsar observations of one-to-two decades. For this reason the
\pt is often ignored. The inclusion of the \pt, and extra information that can
be gathered from it (e.g.\ pulsar distance measurements \citep{Sazhin:1978,
Lommen:2001a, Jenet:2004, Yardley:2010, Corbin:2010, Lee:2011}, source distance
via parallax \citep{Deng:2011} or studying a signal at different times in its
evolution \citep{Mingarelli:2012}), has been studied with regard to continuous
\gw signals, but not for burst sources.

However, pulsar timing arrays (PTAs) consist of many pulsars. This brings the
possibility that among pairs, or more, of pulsars there might be fortuitous
lines-of-sight where the time delay between \pts in particular pulsars is small
enough to be within an observational data set. This is the possibility we
explore in Section~\ref{sec:timedelay} of this paper. 

In a standard burst search using just the \ets all pulsars in the array can be
coherently analysed giving the maximum possible sensitivity, but temporally it
is limited to signals that are passing the Earth now (or at least to within the
decade or so over which the PTA is observing). Looking for signals in the
\pt will not give as good sensitivity as that from a coherent \et search,
because only two (or a few) pulsars can be coherently combined, but each pulsar
pair will provide a different time baseline in which to search. These signals
would otherwise not be observed {\it at all} in an \et-only search. So, given
the temporal coverage it provides it could still be a worthwhile area to search.
We will explore this more in Section~\ref{sec:search}.

\section{Inter-pulsar time delays}\label{sec:timedelay}
\begin{figure}
\begin{tabular}{c}
\includegraphics[width=84mm]{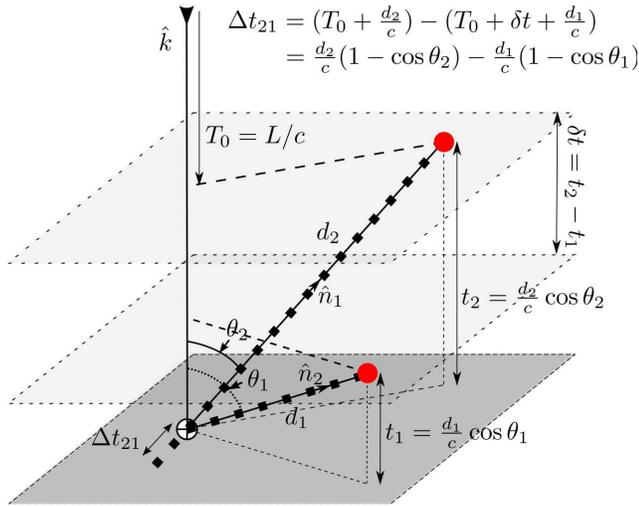}
\end{tabular}
\caption{A schematic diagram showing the geometry of two pulsars, a passing
plane \gw and the Earth $\oplus$. From top to bottom the planes correspond to
the \gw intersecting the first pulsar, then the second pulsar and then the
Earth. The time delay between the \pts is given as $\Delta t_{21}$.}
\label{fig:pulsar_gws}
\end{figure}

\begin{figure}
\begin{tabular}{c}
\includegraphics[width=84mm]{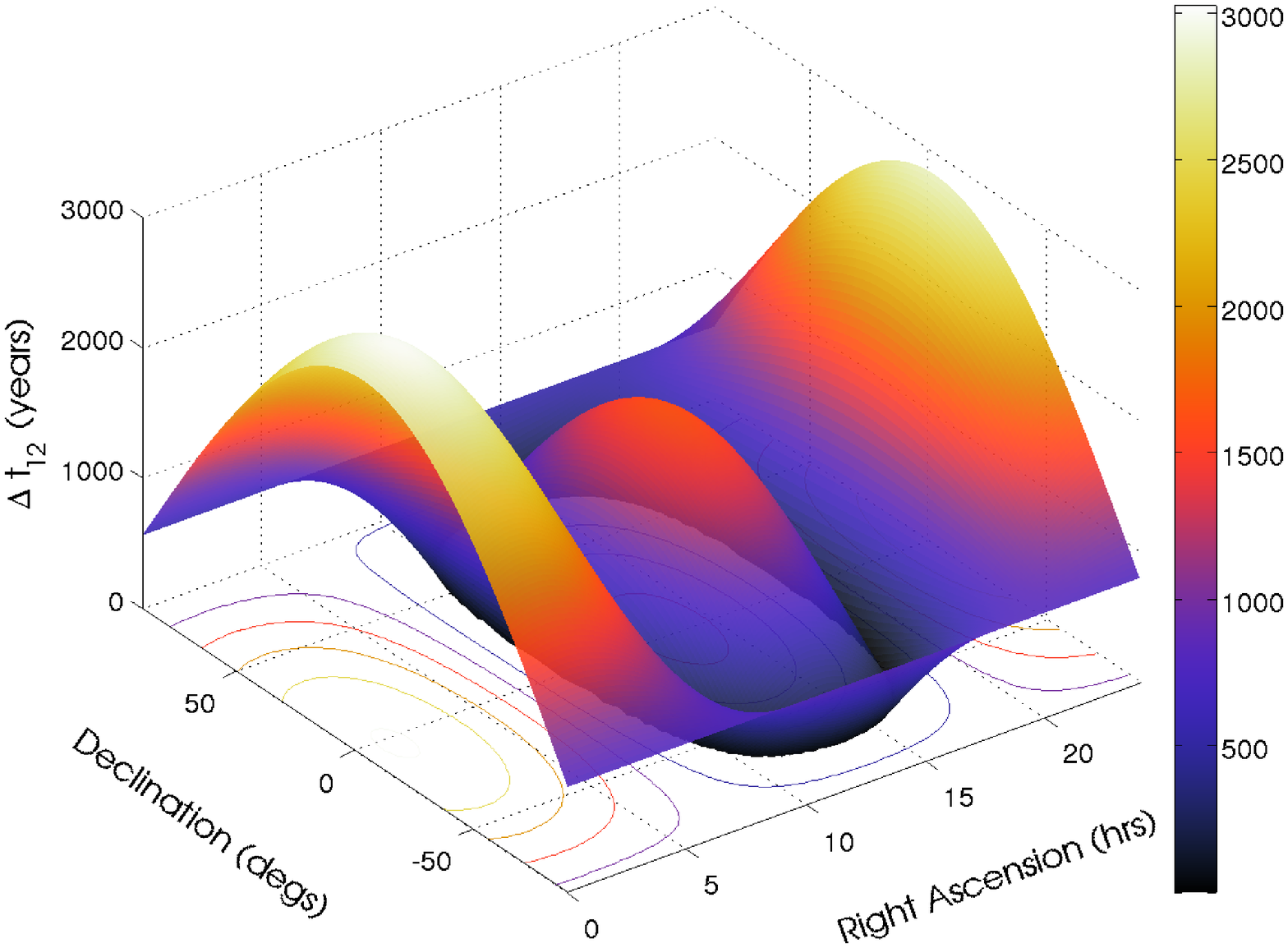}
\end{tabular}
\caption{The time delay between the \pts from Equation~\ref{eq:tdiff}
as a function of source sky position for a pulsar pair consisting of
J1455\textminus3330 (\ra{14}{55}{47}{9}, \dec{-33}{30}{46}{3} and distance of
0.74\,kpc) and J2129\textminus5721 (\ra{21}{29}{22}{7}, \dec{-57}{21}{14}{1}
and a distance of 0.53\,kpc).}
\label{fig:time_delay_example}
\end{figure}

\begin{figure*}
\subfigure[][]{%
\label{fig:td_skymap-a}%
\includegraphics[width=120mm]{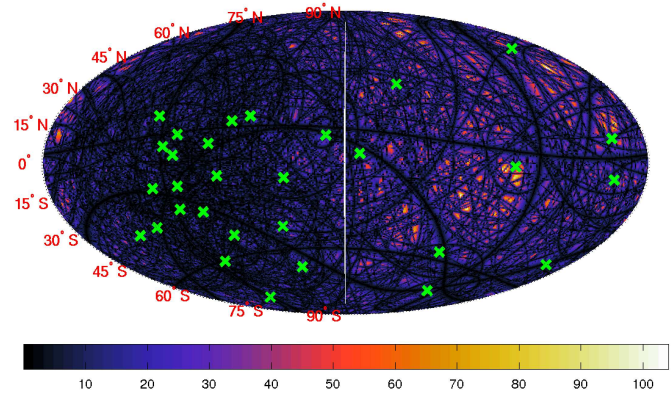} } \\
\subfigure[][]{%
\label{fig:td_skymap-b}%
\includegraphics[width=120mm]{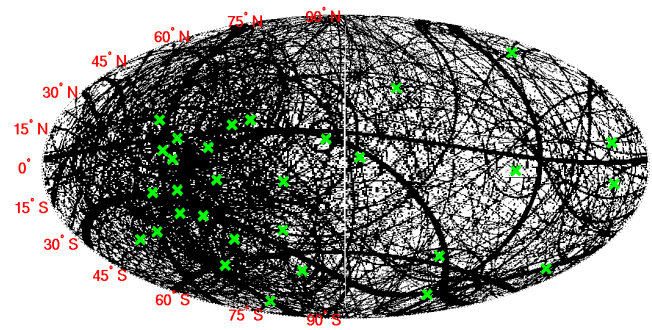} } \\
\subfigure[][]{%
\label{fig:td_skymap-c}%
\includegraphics[width=120mm]{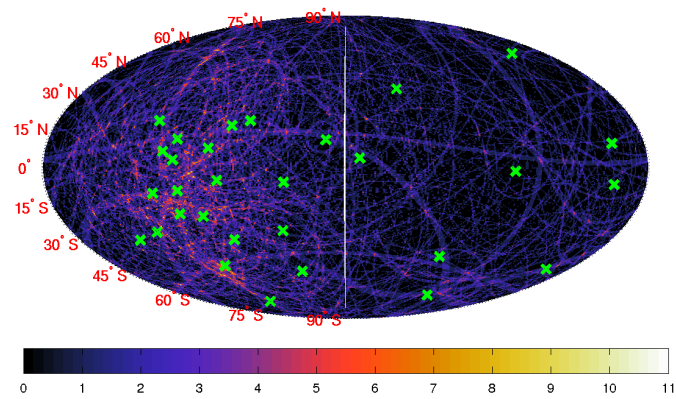} }
\caption{\subref{fig:td_skymap-a} A Hammer projection sky map giving the minimum
time delay between \pts over all IPTA pulsar pairs for a source located at each
position in the sky. The colour range gives the time delay in years.
\subref{fig:td_skymap-b} A map showing only those parts of the sky (black) where
a source will produce a signal separated by less than 10 years within at least
one pulsar pair. \subref{fig:td_skymap-c} A map showing the number of pulsar
pairs for which a source at that sky position would produce a \pt time delay of
less than 10 years. All maps show crosses for the locations of the
IPTA pulsars.}
\label{fig:td_skymap}
\end{figure*}

In this section we will consider the 30 pulsars of the IPTA as given in
\citet{Finn:2010}. Here we look at the minimum time delay between the \pts for
all pairs of pulsars in the array for sources located across the sky. For a
pulsar $i$ with right ascension and declination $\alpha_i$ and $\delta_i$, 
distance $d_i$, and a source position $\alpha_s$ and $\delta_s$, with a
distance $L$ emitting at $t_0$, the time of arrival of the \pt at the Earth will
be
\begin{equation}\label{eq:tarr}
t_i = t_0 + L/c + (1-\cos{\theta_i})d_i/c,
\end{equation}
where the angular separation between the source and pulsar is
\begin{equation}\label{eq:angsep}
\theta_i = \arccos{\{ \cos{(a)}\cos{(b)} +
\sin{(a)}\sin{(b)}\cos{(|\alpha_i - \alpha_s|)}\} },
\end{equation} 
with $a = \pi/2 - \delta_i$ and $b = \pi/2 - \delta_s$. So, for a pair of
pulsars $i$ and $j$ the difference between the time of arrival of the \pt for a
source at a given sky position is
\begin{equation}\label{eq:tdiff}
\Delta{}t_{ij} = |t_i - t_j| = |(1-\cos{\theta_i})d_i -
(1-\cos{\theta_j})d_j|/c.
\end{equation}
A schematic of this set-up for a pair of pulsars is shown in
Figure~\ref{fig:pulsar_gws}. It is important to note that the delay $\delta{}t$
in the figure is not the time delay between the \pts, because the information
that the \gw has influenced the pulsar still has to travel to Earth encoded in
the electromagnetic pulses. Thus the real time difference is given by
$\Delta{}t_{21}$. Figure~\ref{fig:time_delay_example} shows how the time delay,
$\Delta{}t_{21}$, between the times-of-arrival of the \pts observed in pulsar
residuals at Earth for a PTA consisting of a pair of pulsars
(J1455\textminus3330 with \ra{14}{55}{47}{9}, \dec{-33}{30}{46}{3} and distance
of 0.74\,kpc, and J2129\textminus5721 with \ra{21}{29}{22}{7},
\dec{-57}{21}{14}{1} and a distance of 0.53\,kpc) changes for sources located
across the whole sky. It can be seen that the minimum in the delay forms a ring
on the sky. The overlap of many such rings can be seen later in
Figure~\ref{fig:td_skymap}.

\begin{figure}
\begin{tabular}{c}
\includegraphics[width=84mm]{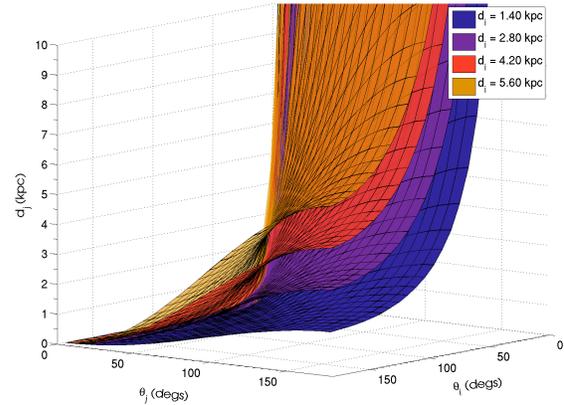}
\end{tabular}
\caption{The sheets represent regions in the $\theta_i, \theta_j, d_i$ and $d_j$
parameter space for a pair of pulsars that fulfil the criterion from
Equation~\ref{eq:tdiff} that $\Delta{}t_{ij} < 10$\,years.}
\label{fig:time_delay_constraint}
\end{figure}

If we choose a constraint such that in Equation~\ref{eq:tdiff}
$\Delta{}t_{ij} < 10$\,years then Figure~\ref{fig:time_delay_constraint} shows
the valid areas of the $\theta_i, \theta_j, d_i$ and $d_j$ parameter space that
fulfil this criterion. It can be seen that in the extreme cases where one pulsar
is closely aligned on the sky with the source, and the other is on the opposite
side of the sky to the source (e.g.\ $\theta_{i,j} \approx 0\degr$ and
$\theta_{j,i} \approx 180\degr$) then the pulsar distance ratio must be very
large to fulfil the criterion.

In this paper we will assume a time delay between \pts of less than 10 years is
reasonable for searches on current, or near future, sets of data. Using a grid
of source sky positions and all the pulsars in the IPTA, we have calculated the
minimum value of $\Delta t_{21}$ in years for all pulsar pairs, the sky area for
which this delay is less than 10 years and the total number of pulsar pairs for
which this delay is less than 10 years. These are shown as Hammer projections
onto the sky in Figure~\ref{fig:td_skymap}. Obviously for longer spans of data
longer time delays can be contemplated and more pulsar pairs will be usable.
Figure~\ref{fig:td_skymap-b} shows the sky area available if we exclude all sky
positions for which no \pt time delay is less than 10 years. This shows that 67
per cent of the sky will contain at least one pair of pulsars with a time delay
between \pts of less than 10 years. Residuals for any such pair could then be
cross correlated with the appropriate delay applied for the sky position.

Figure~\ref{fig:td_skymap-c} shows the total number of pulsar pairs that would
have time delays of less than 10 years for each sky position of the \gw source.
We see that there are some small patches of the sky where up to 11 pulsar pairs
are usable. However, it is unlikely that multiple pairs of pulsars would have
the same individual signal in them for a given sky position. Unfortunately this
means that generally coherent analyses between multiple pairs are not viable,
but having multiple pairs gives you a higher effective temporal observation span
(as discussed in Section~\ref{sec:search}). There are sky locations for which
more than two IPTA pulsars are aligned as such to fulfil the time delay
criterion, but the sky area covered is general small. Figure~\ref{fig:allpta-a}
shows that for the IPTA we find 5.6 per cent of the sky for which the \pt for
three pulsars are within 10 years of each other, 0.3 per cent for which there
are four pulsars and 0.005 per cent for which there are the maximum of five
pulsars.

\subsection{Expanding the PTA}
\begin{figure*}
\subfigure[][]{%
\label{fig:allpta-a}%
\includegraphics[width=120mm]{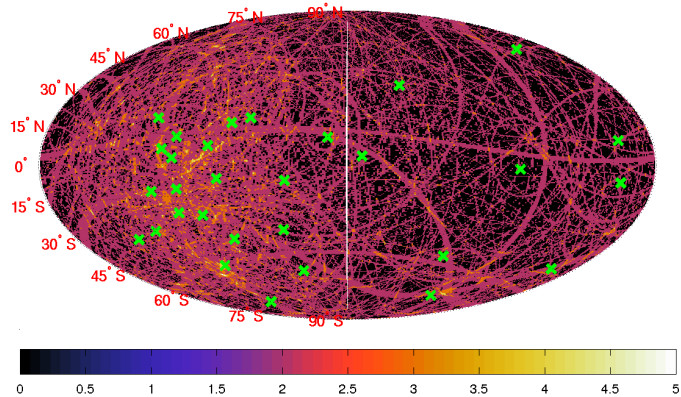} } \\
\subfigure[][]{%
\label{fig:allpta-b}%
\includegraphics[width=120mm]{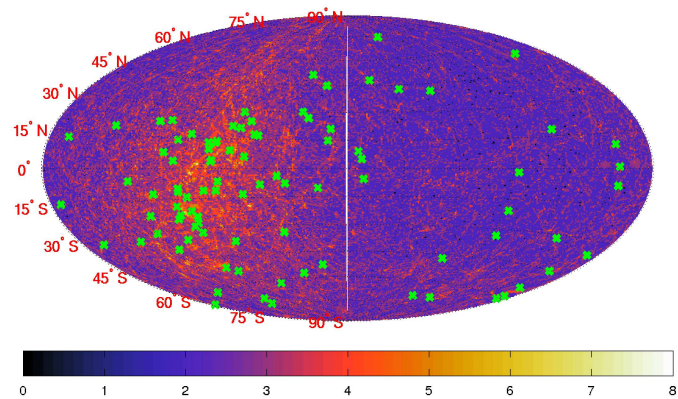} }
\caption{\subref{fig:allpta-a} The number of pulsars with \pts separated by less
that 10 years for given source sky locations using the ITPA.
\subref{fig:allpta-b} The number of pulsars with \pts separated by less that 10
years for given source sky locations using a fake PTA of 91 pulsars.}
\label{fig:allpta}
\end{figure*}

In the future when more pulsars are added to an array more sky areas with
multiple pulsar overlap (i.e.\ $> 2$) may be available, in particular with the
addition of many pulsars within the same globular cluster, which are physically
separated by short (of order years) delays. An example of how co-located pulsars
in globular clusters can be used in \gw detection is given by
\citet{Jenet:2005b}. As a first step at testing how increased numbers of
pulsars improve prospects, we have created a fake array by taking all
millisecond pulsars currently in the Australian Telescope National Facility
(ATNF) pulsar catalogue \citep{Manchester:2005} with frequencies greater than
50\,Hz, but not associated with a globular cluster, leaving 90 pulsars. These
are just used as an example of what happens when adding more pulsars, however it
is not known, or expected, that it will be possible for some, or all, of these
specific pulsars to be included in a future array. With this array, and assuming
a well-known distance to these pulsars, the source locations for which at least
a pair of pulsars has \pts within 10 years show over 99.8 per cent sky
coverage. We find that the number of pulsars for which the \pts are separated by
less than 10 years can be up to 8, although this is for one unique sky position
(\ra{18}{32}{23}{0} and \dec{01}{34}{44}{21}) (given a sky pixel of area $\sim
0.7$ square degrees). The overall number of pulsars with \pts within 10 years
for all source locations across the sky can be seen in
Figure~\ref{fig:allpta-b}. In terms of sky area there is 42 per cent of the sky
for which 3 pulsars have \pts within 10 years, 7 per cent with 4 pulsars, $\sim
1$ per cent with 5 pulsars, $\sim 0.08$ per cent with 6 pulsars and $\sim 0.01$
per cent with 7 pulsars. More pulsars in the array can therefore increase the
sensitivity for certain sky areas. It should be noted that the sky coverage may
be biased by selection effects of observable stable, well-timed, pulsars.

\subsection{Distance uncertainties}
The initial discussions assume that we know the distances to all the pulsars
perfectly, but in reality there will be uncertainties on these distances.
Currently these are optimistically of order 10 per cent, but by the time of the
Square Kilometre Array (SKA) errors may be reduced to less than 1 per cent for
the vast majority of IPTA pulsars, with some of the closest known to maybe a few
tenths of a percent \citep{Smits:2011}. This uncertainty in the distances means
there could be some pulsar pairs that actually lie outside of the required \pt
delay criterion. Conversely, accurate distance estimates could move some pulsar
pairs to within the \pt delay criterion. On average it would be expected that as
many pairs move within the criterion as move out of it, but unless the distance
uncertainties are included in a search it is not known which move in or out.
This means that in reality the area covered in Figure~\ref{fig:td_skymap-b}
would look different, although the total area covered would be approximately the
same.

The effect of this means that for a real search the whole sky would have to be
covered and the uncertainties in the distances to each pulsar taken into
account. So, for each sky position every pulsar pair would have to be tested
with the distances of each varied within the uncertainty range.

\section{Search comparisons}\label{sec:search}
Many methods have been developed to detect, and estimate parameters, for
modelled and unmodelled bursts of \gws in ground-based detectors
\citep[e.g.\ a selection includes][]{Clark:2007, Klimenko:2008, Abbott:2009,
Searle:2009, Sutton:2010}, space-based detectors \citep{Babak:2010} and PTAs
\citep[e.g.][]{Finn:2010, Haasteren:2010}. It would be relatively
straightforward to use similar methods in a search given the situation that we
have presented. So, here rather than define another method we will compare
aspects of a \pt-only search with a more standard \et-only burst search for a
simplified IPTA. The aspects we compare are their sensitivities, their effective
observation times, and from these, the estimated event rates required to give a
detection.

For this we will compare the signal-to-noise ratio that could be
recovered for a single burst source detected in the \ets of the IPTA with that
which could be recovered for a \pt-only search. We will make the simplifying
assumption that all 30 pulsars within the array have residual noise that
is white and Gaussian with standard deviation of 100\,ns, all data spans the
same 20 years period and they are sampled every 30 days. We will require that
to be {\it detected} a signal must have a signal-to-noise ratio ($\rho$) above a
threshold of 10. We will also assume that all the pulsar distances are precisely
known rather than trying to assess the effects searching over some distance
uncertainty has on the threshold required for detection\footnote{In reality the
signal-to-noise ratio threshold at which to set a detection would be lower for
the \et search than the \pt search. This is because the background level (e.g.\
false alarms due to noise) for a signal being observed would be lower due to
more detectors being used and the smaller parameter space from not having to
search over pulsar distances uncertainties. Therefore, the horizon distance for
\et searches could be increased by a small factor of the order of 1.25 reducing
required rates by a factor of $\sim 2$.}.

\subsection{The signal model}\label{sec:signal}
In these tests we assume a simple signal model of a sine-Gaussian burst. This is
defined in the timing residuals of a pulsar (indexed by $i$) by
\begin{eqnarray}\label{eq:fullresidual}
r_i(T) & = &
A_+F_+(\psi,\hat{\bm{k}},\hat{\bm{n}}_i)\Big[\cos{(\omega t_i^e+\phi_0)}
\exp{\Big(-\frac{(t_i^e)^2}{2\tau^2}
\Big)} \nonumber \\
 & & - \cos{(\omega
t_i^p+\phi_0)}\exp{\Big(-\frac{(t_i^p)^2}{2\tau^2}\Big)}\Big] + \nonumber \\
&  &
A_{\times}F_{\times}(\psi,\hat{\bm{k}},\hat{\bm{n}}_i)\Big[\sin{(\omega
t_i^e+\phi_0)}
\exp{\Big(-\frac{(t_i^e)^2}{
2\tau^2 }\Big)} \nonumber \\
 & & - \sin{(\omega t_i^p+\phi_0)}\exp{\Big(-\frac{(t_i^p)^2}{2\tau^2}
\Big)}\Big],
\end{eqnarray}
where $A_+ = A(1+\cos{}^2\iota)/2$ and $A_{\times} = A\cos{\iota}$ for an
amplitude $A$ and source inclination angle $\iota$, $\tau$ is the
Gaussian width, $\omega$ is the angular frequency, $\phi_0$ is the phase at
the midpoint of the burst,  and the definitions of the antenna patterns
$F_{+/\times}$ are described in Appendix~\ref{sec:antenna}. The times are
defined so that the \pt is described by the terms containing $t^p = T - t_i$,
where $T$ is the pulsar proper time at the solar system barycentre and $t_i$ is
time at the observed midpoint of the burst in the \pt (as seen for a particular
pulsar), and therefore the \et time is described by the terms containing $t^e =
T - t_i + (1+\hat{\bm{k}}\cdot\hat{\bm{n}}_i)d_i/c$.  This could be expanded to
a physical
source model such as the parabolic encounter of two supermassive black holes
used in \citet{Finn:2010} (which is qualitatively similar to a sine-Gaussian),
or a more generic burst model as employed in ground-based detector searches.

\subsection{Event rates for detection}
It is useful to try and compare the source event rates that would be required
to detect a signal in the \ets of all pulsars in the IPTA to that which would be
required to detect a signal in the \pts of pairs of pulsars in the IPTA. We have
made some general estimates of this for the sine-Gaussian signal described in
Section~\ref{sec:signal}, with
fixed parameters (following the definition in Equation~\ref{eq:fullresidual}):
$A=500$\,ns, $\omega = 2\pi \times 4$\,yr$^{-1}$, $\psi = 0$\,rads, $\phi_0 =
2.0$\,rads, $\tau = 100$\,days and with the burst centred on the midpoint of
the observations.

We define the signal-to-noise ratio to a given signal, $r$, for a set of $N$
pulsars as
\begin{equation}\label{eq:snr}
\rho = \left( \sum_{i=1}^N \sum_{j=1}^{n_i} \frac{r_j^2}{\sigma_i^2}
\right)^{1/2},
\end{equation}
where $n_i$ is the number of residual data points, and $\sigma_i$ is the noise
standard deviation, for the $i^{\rmn th}$ pulsar (working in the time domain).

\subsubsection{Search sensitivity and range}
\begin{figure*}
\subfigure[][]{%
\label{fig:skysnr-a}%
\includegraphics[width=105mm]{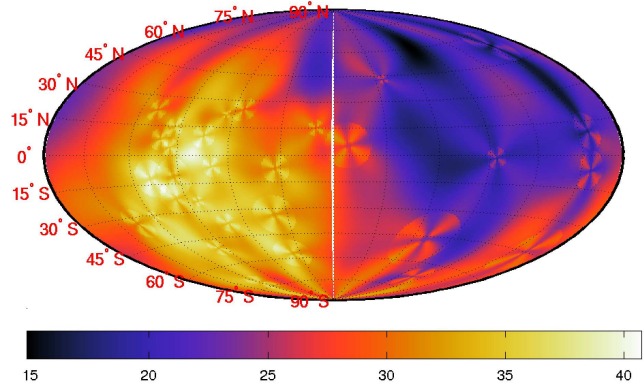} } \\
\subfigure[][]{%
\label{fig:skysnr-b}%
\includegraphics[width=105mm]{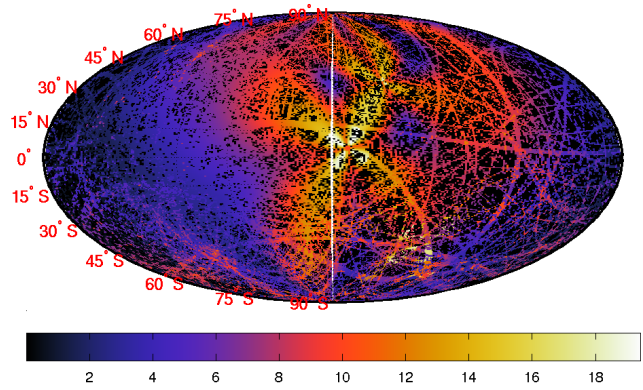} } \\
\subfigure[][]{%
\label{fig:skysnr-c}%
\includegraphics[width=105mm]{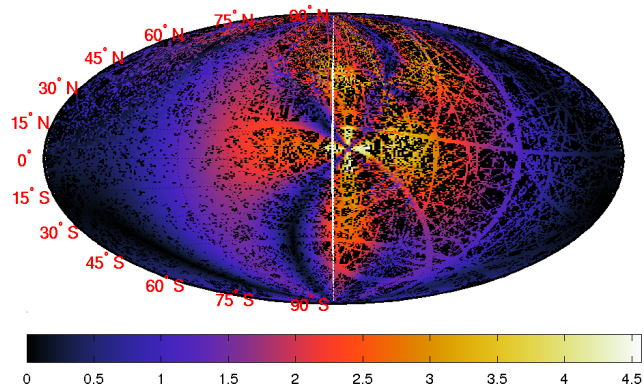} }
\caption{Hammer projection sky maps giving the signal-to-noise ratios for a
source located at different sky locations for the IPTA observations with:
\subref{fig:skysnr-a} the signal appearing in the \ets of all pulsars
for an optimally oriented source; \subref{fig:skysnr-b} the signal appearing
in the \pts for pairs (or multiples) of pulsars for an optimally oriented source
and using the loudest pair for a given sky location; and, \subref{fig:skysnr-c}
the signal appearing in the \pts for pairs (or multiples) of pulsars for the
worst source orientation and using the quietest pair for a given sky location.}
\label{fig:skysnr}
\end{figure*}

Using Equation~\ref{eq:snr} we have calculated the signal-to-noise ratio for the
sine-Gaussian source defined above at each point in the sky for the IPTA
pulsars. Figure~\ref{fig:skysnr-a} shows the signal-to-noise ratio that this
source (if optimally oriented with $\cos{\iota} = \pm 1$) could be observed at
if located across the sky and seen in the \ets for the entire IPTA. The
flower-like patterns in the response across the sky come from the antenna
patterns of pairs of pulsars as seen in Figures~\ref{fig:antenna1} and
\ref{fig:antenna2}. The range of signal-to-noise ratios for this particular
source over the sky was between 40.7 and 14.8, whereas if the the worst case
source orientation were chosen ($\cos{\iota} = 0$) then the signal-to-noise
ratio was between 10.0 and 2.5. 

Figure~\ref{fig:skysnr-b} shows the {\it maximum} signal-to-noise ratio that
this source (if optimally oriented with $\cos{\iota} = \pm 1$) could be observed
at if located across the sky and seen in the \pts of pairs (or multiples when
applicable) of IPTA pulsars. We define regions of the sky with no pulsar
pairs \pts separated by less than 10 years as having zero signal-to-noise i.e.\
we assume that we could not see, or would ignore as potential noise, a \pt
signal only observed in one pulsar. In this case what we mean by {\it maximum}
is that for locations for which multiple pulsar pairs could contain signals we
have taken the signal-to-noise ratio of the loudest pair. Over the sky area that
could be observed ($\sim 67$ per cent of the sky) the range of signal-to-noise
ratios for the {\it maximum} and optimally oriented case was between 19.4 and
0.4, whereas for the worst case orientation (and with the quietest pulsar pair
chosen) the ranges was between 4.6 and $7\ee{-4}$ (see
Figure~\ref{fig:skysnr-c}).

The ratio of signal-to-noise ratios between an \et search (using all 30
pulsars) and a \pt search (using only a pair of pulsars for each sky position)
agree well with the simple calculation of $\sqrt{30/2}\approx 4$.

From Figure~\ref{fig:skysnr} we can calculate a sky-averaged signal to noise
ratios, $\langle \rho \rangle$, for a detectable source. For the example case
above of using the \ets for the search and an optimally oriented source
$\langle \rho \rangle = 26.7$. We know that the signal amplitude that led to
this was $A = 500$\,ns, so assuming that we want a source to have $\langle \rho
\rangle \ge 10$ for detection we can scale the signal amplitude
appropriately giving an effective sky averaged amplitude required for detection
of
\begin{equation}\label{eq:avamp}
\langle A \rangle \ge 500 \times \left( \frac{10}{26.7} \right)\,{\rmn{ns}} =
187\,{\rmn{ns}}. 
\end{equation}
To set a scale (which will effect the absolute values of our event rates
estimated below, but will not effect the relative ratios between them), we will
say that a source at a distance of 1\,Gpc produces an amplitude in the pulsar
residuals of 100\,ns. The observed amplitude is directly proportional to the
source distance $D$, so for detection we require that
\begin{equation}\label{eq:range}
\left( \frac{D}{1\,{\rmn{Gpc}}} \right) \le \left( \frac{100\,\rmn{ns}}{\langle
A \rangle\,\rmn{ns}} \right),
\end{equation}
or more generically
\begin{equation}
\left( \frac{D}{1\,{\rmn{Gpc}}} \right) \le \left( \frac{100\,\rmn{ns}}{
A\,\rmn{ns}} \right) \left(\frac{ \langle \rho \rangle_A }{
\langle \rho \rangle_{\rmn{thresh}}}\right),
\end{equation}
where $A$ is the observed signal amplitude that gives a sky-averaged
signal-to-noise ratio of $\langle \rho \rangle_A$, and $\langle \rho
\rangle_{\rmn{thresh}}$ is the signal-to-noise ratio threshold for detection
(so given the case in Equation~\ref{eq:avamp} $A = 500$\,ns, $\langle \rho
\rangle_A = 26.7$ and $\langle \rho \rangle_{\rmn{thresh}} = 10$). This gives us
a sky-averaged range, or horizon distance \citep[similar to that used as a
figure of merit for compact binary coalescences in ground-based \gw detectors
e.g.][]{Abbott:2005}, as a function of the observed amplitude.

From Figure~\ref{fig:skysnr} we can calculate that for the \et search the best
and worst case sky-averaged signal-to-noise ratios, $\langle \rho \rangle$ are
26.7 and 6.4 respectively, giving horizon distances of 0.53 and 0.13 Gpc. For
the \pt search the best and worst case sky-averaged signal-to-noise ratios,
$\langle \rho \rangle$ are 6.6 and 1.3 respectively, giving horizon distances of
0.13 and 0.03 Gpc.

\subsubsection{Effective observation times}
For the \et search each pulsar observes the whole sky, so the effective total
observation time, $T_{\rmn{tot}}$, is just the span of the residual observations
$T_{\rmn{tot},ET} = T_{\rmn{res}}$, but for the \pt search the calculation is
more complex. For a pulsar pair the overlapping observation time that
could contain a signal (assuming that the signal width is small compared to the
data span) will be $T_{\rmn{res}} - \Delta t$, where $\Delta t$ is the maximum
delay between \pts. Also, whereas in the \et search each pulsar observes the
whole sky, in the \pt search each pulsar pair will only have a fractional sky
coverage that gives \pts within $\Delta t$ (the ring on the sky in
Figure~\ref{fig:time_delay_example}). So, for a particular pulsar pair the
observation time factored by the sky coverage will be
\begin{equation}
T_{{\rmn{obs}},i} = (T_{\rmn{res}} - \Delta{}t)\times f_i,
\end{equation}
where $f_i$ is the fractional sky coverage for that pair. The effective total
observation time will therefore be $T_{\rmn{tot},PT} = \sum_i^N
T_{{\rmn{obs}},i}$, where $N$ is the number of pulsar pairs. As we are assuming
$T_{\rmn{res}}$ and $\Delta{}t$ are the same for all pulsar pairs this becomes
\begin{equation}
T_{\rmn{tot},PT} = (T_{\rmn{res}} - \Delta{}t) \times \sum_i^N f_i.
\end{equation}
So, for the \pt search the lack of sky coverage for individual pulsar pairs can
be compensated for by the effective increase in the temporal coverage i.e.\ each
pulsar pair sees a different time epoch, but can only see that epoch for a small
portion of the sky.

For $T_{\rmn{res}}=20$ years and $\Delta{}t = 10$ years we find that for all
IPTA pulsar pairs the sum of their fractional sky coverage is $\sum_{i=1}^{435}
f_i = 1.15$. This means that $T_{\rmn{tot},PT} = 20$ years and $T_{\rmn{tot},ET}
= 11.5$, which are comparable. In the next section we discuss what this
indicates in terms of event rates.

\subsubsection{Source event rates}
If we assume that the rate of events that would produce an observed signal
amplitude of 100\,ns if at a distance of 1\,Gpc, $R$ (per unit volume per unit
time), is the same throughout the Universe we can approximate the observed
number of events, $O$, as
\begin{equation}\label{eq:rate}
O = R \times \frac{4}{3}\pi D^3 \times T_{\rmn{tot}}.
\end{equation}

\begin{figure}
\includegraphics[width=84mm]{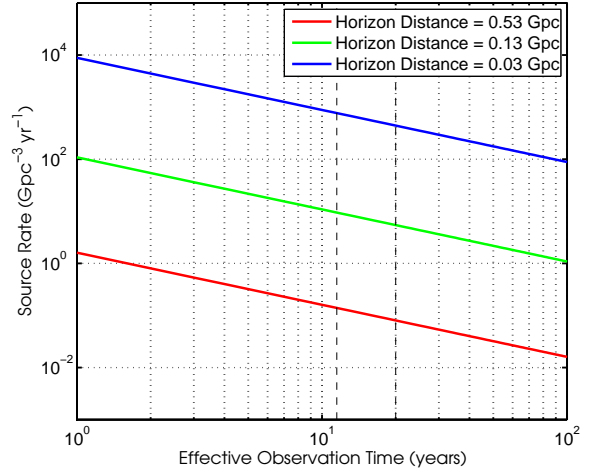}
\caption{The source event rate as a function of effective observation times
required to observed one event for a variety of detector horizon distances. The
vertical dashed lines show the effective observation times for a search using
the \pts (left) and \ets (right) with the IPTA pulsars.}
\label{fig:tobs_vs_rate}
\end{figure}

Using the horizon distances calculated above we can use Equation~\ref{eq:rate}
to calculate the rate of these events (i.e.\ the rate of sources that would
produce a 100\,ns amplitude signal if at 1\,Gpc) against total observation time
if we want to be able to observe one signal. This is shown in
Figure~\ref{fig:tobs_vs_rate} along with lines corresponding to the effective
observation times of an \et search and a \pt search.
We find that for the \et search, with an effective observation time of
$T_{\rmn{tot}} = 20$ years, source event rates of $\sim
0.08$\,Gpc$^{-3}$\,yr$^{-1}$ and $\sim 5.5$\,Gpc$^{-3}$\,yr$^{-1}$ would be
required to observe one event assuming all sources are either in the best
case or worst case orientations respectively. For a search using the \pts, with
an effective observation time of $T_{\rmn{tot}} = 11.5$ years, source event
rates of $\sim 9.6$\,Gpc$^{-3}$\,yr$^{-1}$ and $\sim 790$\,Gpc$^{-3}$\,yr$^{-1}$
would be required to observe one event with the best case and worst case
orientations respectively. This suggests that many sources should be observed in
an \et search if you were to observed one in the \pt search, or conversely if
you observed one event in an \et search you would expect it to be very unlikely
that a \pt search would also produce an observed signal.

\subsubsection{An expanded array}
Using the example of the IPTA above we can scale values to see how \et searches
and \pt searches would compare for a larger array of $N$ pulsars given a fixed
source event rate R. For an \et search, given that the observation time is
fixed, the number of events that will be observed is dependent on the observing
volume, which itself depends in the horizon distance, which corresponds to the
observed sky-averaged signal-to-noise ratio. The signal-to-noise ratio increases
with the square root of the number of pulsars in the array, so overall the
observed number of events becomes
\begin{equation}\label{eq:etrate}
O_{\rmn{\it ET}} = R \times \frac{4}{3}\pi \left(D_{
ET,{\rmn{IPTA}}}\left(\frac{N}{30}\right)^{1/2}\right)^3 \times
T_{\rmn{tot},ET}
\end{equation}
where $D_{ET,{\rmn{IPTA}}}$ and $T_{ET,{\rmn{tot}}}$ are the horizon distance
and effective observation times for the \et search using the IPTA above. For the
\pt search the signal-to-noise ratio, and hence horizon distance and observed
volume, is fixed (assuming that we still only use pairs of pulsars, rather than
more), but the effective total observation time changes. If each pulsar pair
sees on average a fraction $\bar{f}$ of the sky, and the number of pulsar pairs
given $N$ pulsars is $N(N-1)/2$, then the effective observation time is
proportional to
$\bar{f}N(N-1)/2$, and the observed number of events becomes
\begin{equation}\label{eq:ptrate}
O_{\rmn{\it PT}} = R \times \frac{4}{3}\pi D_{PT,{\rmn{IPTA}}}^3 \times
T_{PT}\bar{f}N(N-1)/2,
\end{equation}
where $D_{PT,{\rmn{IPTA}}}$ is the horizon distance for the \pt search using
the IPTA above and $T_{PT} = T_{\rmn{res}} - \Delta{}t = 10$ years as above.
It can be seen that Equation~\ref{eq:etrate} scales as $N^{3/2}$ whereas
Equation~\ref{eq:ptrate} scales as $N^2$, so for a fixed rate (provided the
sources and event rates are isotropic within the horizon distance) and a large
enough array of pulsars the number of events observable in a \pt search should
overtake that observable in an \et search, i.e.\ the total effective observation
time available will compensate for the smaller range.

\begin{figure}
\includegraphics[width=84mm]{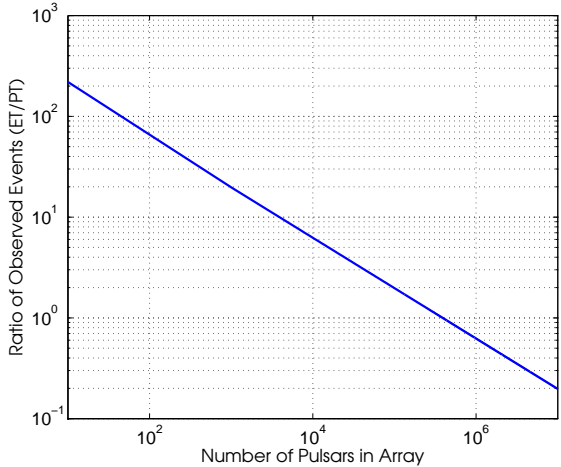}
\caption{The ratio of the number of events observable in an \et search
compared to that observable in an \pt search as a function of the number of
pulsars in a PTA.}
\label{fig:observed_events}
\end{figure}

Using the numbers for the IPTA (i.e.\ assuming a maximum delay between \pts of
10 years) we see that the average sky fraction observed by a pulsar pair is
$\bar{f} = 1.15/435 = 0.0026$. Using the horizon distances for the best case
orientation for the \et and \pt searches Figure~\ref{fig:observed_events} shows
the ratio of the observable number of events in an \et search to a \pt search as
a function of the number of pulsars in an array.
We see that to achieve equality in the number of events observed would require
an array of $\sim 400\,000$ pulsars. This is greater than the total expected
number of active pulsars in the Galaxy \citep{Lorimer:2008}, so is not
achievable.

\section{Summary}
In searches for stochastic or burst sources of gravitational waves in PTA data
the delayed \pt is often ignored as it will generally be incoherent between
residuals from different pulsars. We have looked at whether extra value could
be gained when looking for short duration burst sources by looking for coherent
signals in \pts from pairs, or more, of pulsars for specific source sky
locations. In the era of the Low Frequency Array (LOFAR)/SKA it could be
possible to observe a few thousand millisecond pulsars \citep{Smits:2009}, with
around 100 of the best timed of these being usable as ``arms'' in a PTA
\citep{Kramer:2012}. However, our studies have mainly focused on the
possibilities using the IPTA, containing 30 pulsars.

We have shown that for a PTA timing array there can be significant
areas of the sky containing a source for which the \pts of a signal can appear
separated by relatively short times in pairs of pulsar residuals. In particular
for the IPTA we find that this is the case for 67 per cent of the sky given
maximum delays between \pts in two pulsars of 10 years. To accurately
know which sky locations this is valid for would require precise (sub one per
cent) knowledge of the PTA pulsar distances. This may be possible in the future
with the SKA, but otherwise any search would have to take into account the
distance uncertainties. For arrays of more pulsars the sky coverage becomes
greater, e.g.\ for 90 pulsars almost the whole sky is covered, and there also
become more source sky locations that give \pts observable in more than two
pulsars. 

We have also compared the relative sky-averaged sensitivities of a search using
the \ets, which are coherent between all pulsars in the PTA, with that of a
search that uses the \pts in pairs of pulsars. Considering an optimally
oriented source, and using the pulsar pair with the largest antenna pattern for
a given sky location, we find that the \et search is on average about four times
more sensitive. For the worst case source orientation, and using the pulsar pair
with the smallest antenna pattern we find that the \et search is on average five
times more sensitive. These are very comparable with what might be expected
given that the \et search uses 30 pulsars, whereas the \pt search is generally
just using pairs of pulsar i.e.\ $\sqrt{30/2} \approx 4$. An \et search would
require a signal to occur within the observation window of the residuals,
whereas each \pt search would be looking back in time covering some time window
dependent of the delay between pulsar terms. However, the \et search would
cover the whole sky whereas the searches for \pts in pairs of pulsars would
only cover small parts of the sky. For residual observations covering 20 years
and maximum delays between \pts of less than 10 years we have found that a \pt
search would have an effective observation time of 11.5 years taking into
account the limited sky coverage for each pulsar pair. We have converted these
sensitivities to horizon distances for a putative source, and along with the
effective observation times, used them to estimate the event rates required to
observe one event in the \ets and \pts of a simulated IPTA with 20 years of
residuals. We have found that event rates for comparable sources show that you
would have approximately two orders of magnitude higher chance of observing a
signal in the \ets than the \pts for an {\it all-sky} search. However, this does
not mean that \pt searches are pointless. They could be complementary to \et
searches in helping add more observation time to specific targets in the sky
like large galaxy clusters. We have also assumed that all pulsar residuals have
equivalent noise levels, but it could be that the least noisy pulsars could
dominate the sensitivity and improve the comparisons.

For arrays of larger numbers of pulsars we have seen that the scaling of the
total effective observation time for a \pt search rises faster than the
search volume of the \et search. However, unfortunately the potential observable
number of events in a \pt search does not rise above that for an \et search
until $\sim 400\,000$ pulsars are in the array. Unfortunately such a large array
is a completely unachievable goal. However, in larger arrays much more of the
sky would be covered be cases when more that two pulsars have \pts separated by
less than 10 years, so this would potentially bring that ratio down to a more
hopeful level.

In future work we plan to: investigate a more realistic timing array with
more physical pulsar residuals; test and characterise a search routine; and, if
possible, apply it to real data, or future IPTA mock data
challenges\footnote{\url{http://www.ipta4gw.org/}}. These studies will show
whether the results here are more pessimistic than necessary.

\section*{Acknowledgements}
This work has been funded under a UK Science and Technology Facilities Council
rolling grant. I would like to thank Andrea Lommen and Graham Woan for
discussions that lead to this work and for comments on drafts of this paper.
I thank the referees for many valuable comments on the paper and in particular
for the suggestion to look at comparisons between ``Earth term'' and ``pulsar
term'' searches.

\bibliographystyle{mn2e}
\bibliography{pta_burst_detection}

\appendix

\section{The antenna pattern}\label{sec:antenna}
\begin{figure}\includegraphics[width=84mm]{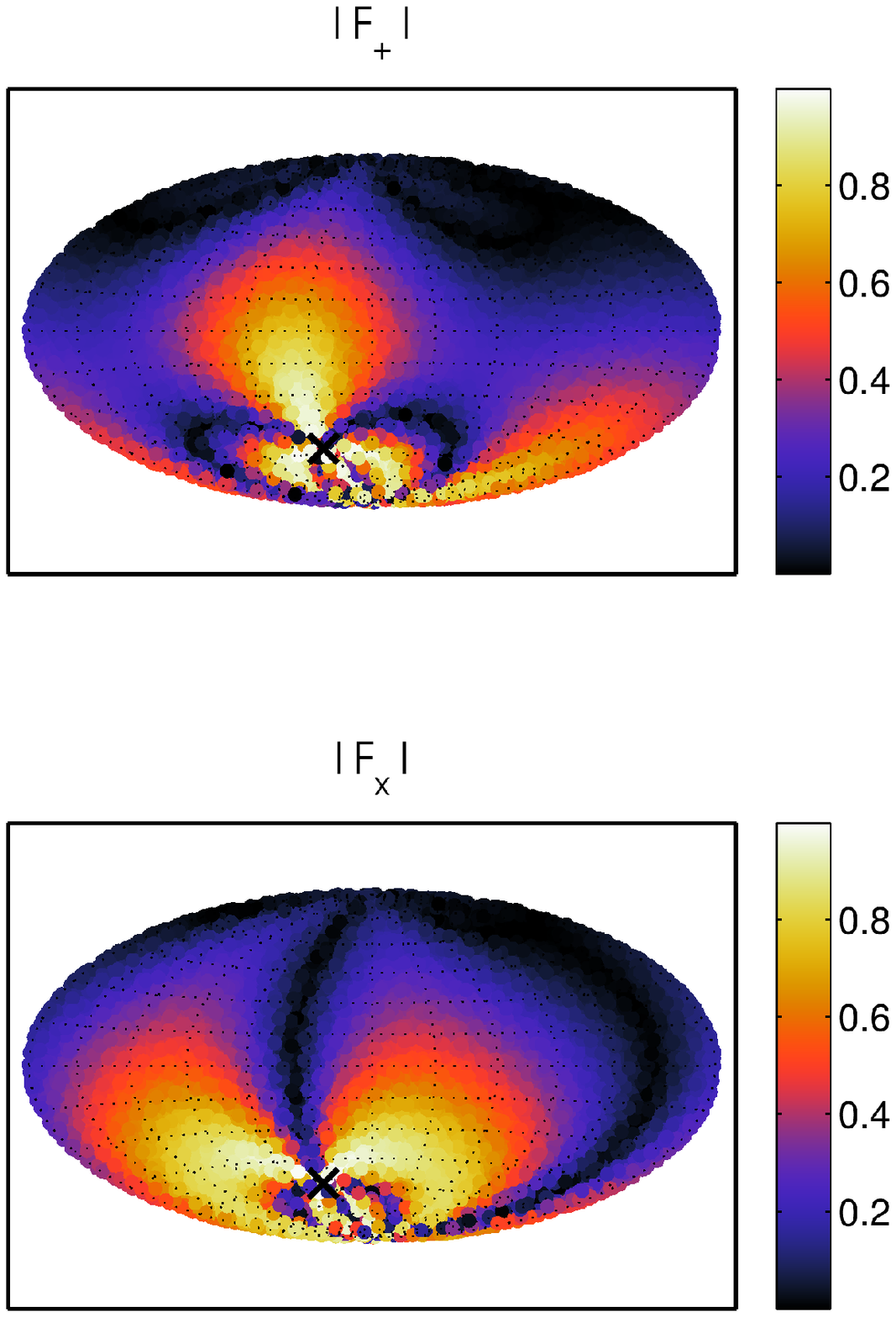}
\caption{The absolute values of the `plus' and `cross' polarisation antenna
pattern for pulsar J2129\textminus5721 (\ra{21}{29}{22}{7},
\dec{-57}{21}{14}{1}) over the whole sky. The pulsar location is marked by the
black cross.}
\label{fig:antenna1}
\end{figure}

\begin{figure}\includegraphics[width=84mm]{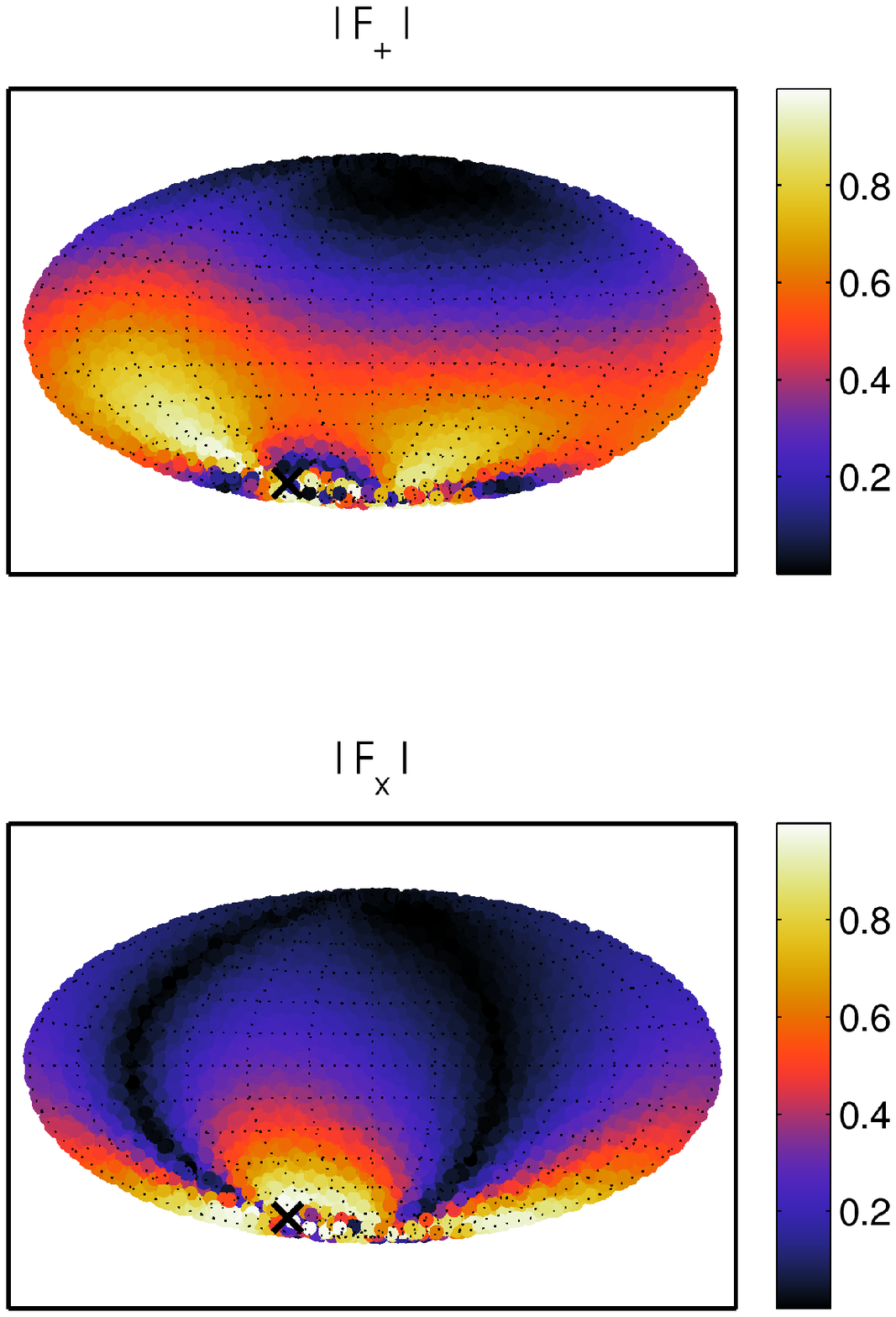}
\caption{The absolute values of the `plus' and `cross' polarisation antenna
pattern for pulsar J1603\textminus7202 (\ra{16}{03}{35}{6},
\dec{-72}{02}{32}{6}) over the whole sky. The pulsar location is marked by the
black cross.}
\label{fig:antenna2}
\end{figure}

Given $\hat{\bm{k}}$ is a unit vector pointing along the wave propagation
direction
from the source to the Earth we can construct the following basis vectors
representing the wave propagation frame,
\begin{eqnarray}
\hat{\bm{k}} & = & (-\sin{\theta_s}\cos{\phi_s}, -\sin{\theta_s}\sin{\phi_s},
-\cos{\theta_s}),\\
\hat{\bm{l}} & = & (\sin{\phi_s}, -\cos{\phi_s}, 0),\\
\hat{\bm{m}} & = & (\cos{\theta_s}\cos{\phi_s}, \cos{\theta_s}\sin{\phi_s},
-\sin{\theta_s}),
\end{eqnarray}
and define the unit vector from the Earth (or more correctly the
solar system barycentre) to the pulsar $\hat{\bm{n}} =
(\sin{\theta_p}\cos{\phi_p},
\sin{\theta_p}\sin{\phi_p}, \cos{\theta_p})$, where $\phi_{s,p}$ give the right
ascensions and $\theta_{s,p} = \pi/2 - \rmn{declination}$ for the source and
pulsar respectively. From these the polarisation basis tensors can be defined by
\begin{eqnarray}
e_+^{ab} & = & l^a l^b - m^a m^b, \\
e_{\times}^{ab} & = & l^a m^b + m^a l^b,
\end{eqnarray}
and, with the inclusion of the polarisation angle $\psi$, by
\begin{eqnarray}
e_+^{ab} & = & \cos{(2\psi)} e_+^{ab} + \sin{(2\psi)} e_{\times}^{ab}, \\
e_{\times}^{ab} & = & -\sin{(2\psi)} e_+^{ab} + \cos{(2\psi)} e_{\times}^{ab}.
\end{eqnarray}
The `plus' and `cross' antenna patterns (e.g.\ see \citealt{Anholm:2009} or
\citealt{Finn:2010}) are therefore given by
\begin{equation}
F_{+/\times} = \frac{1}{2}e_{+/\times}^{ab}
\frac{\hat{\bm{n}}_a\hat{\bm{n}}_b}{1+\hat{\bm{k}}\cdot\hat{\bm{n}}}.
\end{equation}
Examples of these full sky antenna patterns for a couple of pulsars, with
$\psi = 0$, are given in Figures~\ref{fig:antenna1} and \ref{fig:antenna2}.


As we are not in the long wavelength approximation regime pulsar timing
residuals will consist of components from the \et and \pt, such that
\begin{eqnarray}
r(T) & = & (A_+(t^e) - A_+(t^p))F_+(\psi,\hat{\bm{k}},\hat{\bm{n}}) + \nonumber
\\
& & (A_{\times}(t^e) -
A_{\times}(t^p))F_{\times}(\psi,\hat{\bm{k}},\hat{\bm{n}})
\end{eqnarray}
where, as we are dealing with the \pts in this paper, we set the time of the
\pt as $t^p = T - t_i$, where $T$ is the pulsar proper time at the solar system
barycentre and $t_i$ is time at the observed midpoint of the burst in the pulsar
term (as seen for a particular pulsar), and therefore the \et time is
$t^e = T - t_i + (1+\hat{\bm{k}}\cdot\hat{\bm{n}}_i)d_i/c$. 

When creating a coherent signal in two pulsars' residuals the signal model for
the second pulsar will have to be shifted by $\Delta{}t_{ij} =
(1+\hat{\bm{k}}\cdot\hat{\bm{n}}_i)d_i/c -
(1+\hat{\bm{k}}\cdot\hat{\bm{n}}_j)d_j/c$ with respect to
the first pulsar i.e.\ if we have $r_1(T)$ then to coherently combine it
with another dataset that would have to have $r_2(T + \Delta{}t_{12})$, with
$t_i = t_1$ for both models.

\end{document}